\documentclass[aps,prl,twocolumn,superscriptaddress,showpacs,preprintnumbers,amsmath,amssymb,floatfix,longbibliography]{revtex4-2}
    \usepackage{graphicx} 
    \usepackage{dcolumn} 
    \usepackage{bm} 
    \usepackage{color} 
    \usepackage{lipsum} 
    \usepackage{graphicx}
    \usepackage{dcolumn}
    \usepackage{bm}
    \usepackage{siunitx} 
    \usepackage{bbold}
    \usepackage{booktabs}
    \usepackage{subfiles}
    \usepackage{python}
    \usepackage{enumitem}
    \usepackage{lineno}
    \usepackage[usenames,dvipsnames]{xcolor}
    \usepackage[linktocpage, colorlinks=true, linkcolor=blue, citecolor=blue, breaklinks=true, urlcolor=blue]{hyperref}
    \usepackage{tikz}
    \usepackage{scalerel}
    \UseRawInputEncoding
    \usetikzlibrary{svg.path}
    \renewcommand{\vec}[1]{\bm{#1}}

    \usepackage{comment}
    \usepackage{verbatim}
    \definecolor{orcidlogocol}{HTML}{A6CE39}
    \tikzset{
      orcidlogo/.pic={
        \fill[orcidlogocol] svg{M256,128c0,70.7-57.3,128-128,128C57.3,256,0,198.7,0,128C0,57.3,57.3,0,128,0C198.7,0,256,57.3,256,128z};
        \fill[white] svg{M86.3,186.2H70.9V79.1h15.4v48.4V186.2z}
                     svg{M108.9,79.1h41.6c39.6,0,57,28.3,57,53.6c0,27.5-21.5,53.6-56.8,53.6h-41.8V79.1z M124.3,172.4h24.5c34.9,0,42.9-26.5,42.9-39.7c0-21.5-13.7-39.7-43.7-39.7h-23.7V172.4z}
                     svg{M88.7,56.8c0,5.5-4.5,10.1-10.1,10.1c-5.6,0-10.1-4.6-10.1-10.1c0-5.6,4.5-10.1,10.1-10.1C84.2,46.7,88.7,51.3,88.7,56.8z};
      }
    }
    \newcommand\orcid[1]{\href{https://orcid.org/#1}{\mbox{\scalerel*{
    \begin{tikzpicture}[yscale=-1,transform shape]
    \pic{orcidlogo};
    \end{tikzpicture}
    }{|}}}}

\begin{document}
    \setcounter{secnumdepth}{3}
    \title{Active diffusion enhances plankton carbon capture and phycosphere radius}
    
    \author{Maggie Liu\orcid{0009-0003-8727-6885}}
        \affiliation{Department of Physics \& Astronomy,
        		University of Pennsylvania, Philadelphia, PA 19104}
    \author{Arnold J. T. M. Mathijssen\orcid{0000-0002-9577-8928}}
        \email{amaths@upenn.edu}
        \affiliation{Department of Physics \& Astronomy,
        		University of Pennsylvania, Philadelphia, PA 19104}
    \date{\today}

\begin{abstract}
Plankton fix about 40 gigatons of carbon annually, using photosynthesis to convert $\text{CO}_2$ into $\text{O}_2$ and carbohydrates. These solutes are exchanged with the ocean in a diffusive boundary layer around the organism called the phycosphere. Here, we study how organisms can increase their carbon influx and outflux by actively mixing the surrounding fluid. By developing exact analytical expressions validated by stochastic simulations, we determine the enhanced diffusivity of phycosphere particles as a function of mixing activity, and their resulting fluxes and concentration fields. Hence, we find that plankton can significantly increase their uptake and photosynthetic turnover. Moreover, we find that the phycosphere radius is enlarged both by increased metabolism and by increased diffusive transport further from the organism. These results provide new biophysical insights into marine microbial ecology, with important implications for global carbon capture and climate change.
\end{abstract}


\maketitle

\section{Introduction}
\label{sec:intro}

Phytoplankton are autotrophic unicellular organisms that account for about \(40\%\) of all global carbon fixation and around \(70\%\) of the oxygen in our atmosphere through photosynthesis \cite{Hays2005ClimatePlankton, Guasto2012FluidMicroorganisms, Rusconi2015MicrobesFlow, Wheeler2019NotPlankton}.
The fixed carbon produced by these cells enters marine food webs directly and also fuels heterotrophic bacteria through the microbial loop \cite{Azam1983EcologicalRole,Buchan2014MasterRecyclers,Jiao2024TheChange}.

At the single-cell scale, plankton exchange dissolved molecules with the surrounding ocean through the phycosphere, a chemically enriched boundary layer surrounding each cell \cite{Seymour2017ZoomingRelationships,Smriga2016ChemotaxisBacteria,Amin2015InteractionBacteria}. 
At this interface, inorganic carbon and nutrients are captured while oxygen and dissolved organic metabolites are released \cite{Prairie2012BiophysicalReview}.
The resulting chemical gradients can attract marine bacteria, alter encounter and colonization rates, and shape remineralization around individual phytoplankton cells \cite{Stocker2012MarineGradients,Smriga2016ChemotaxisBacteria,Stocker2008RapidPatches,Forget2026ThePhytoplankton,Dow2021HowInteractions}. 
These bacteria-phycosphere interactions are increasingly viewed as a central component of phytoplankton physiology and community function \cite{Amin2015InteractionBacteria,Buchan2014MasterRecyclers,Semple2025EngineeringConsortia}.

At larger scales, phytoplankton-derived organic matter can seed transparent exopolymer particles and marine snow, creating microbial hotspots within the biological carbon pump \cite{Passow2002TransparentEnvironments, Decho2017MicrobialSystems, Kirboe2001MarineBacteria, Boyd2019Multi-facetedOcean, Alldredge1988CharacteristicsSnow, Azam2001SeaMicrocosms, Simon2002MicrobialEcosystems, Smith1992IntenseDissolution}. 
Marine snow aggregates contain steep chemical gradients and active bacterial degradation, so mechanisms that expand phycosphere chemical fields may influence not only cell-scale encounters, but also the early development of aggregate-associated communities \cite{Kirboe2001MarineBacteria,Smith1992IntenseDissolution}.


Classical descriptions of phycosphere transport treat uptake and release as passive molecular diffusion around an absorbing or excreting body \cite{Berg1977PhysicsChemoreception,Munk1952AbsorptionNutrients,PasciakGavis1974TransportLimitation,KarpBoss1996NutrientFluxes,Lindemann2016ScalingAffinity}. 
In this limit, the physical cell radius sets the characteristic flux scales: 
The diffusive influx grows linearly with cell size, while volume-based metabolic demand grows more rapidly.
This mismatch motivates long-standing questions about how larger or more active plankton overcome this influx limitation set by diffusion \cite{Gavis1976MunkGrowth,Breckels2010ModellingCells}.
Many phytoplankton and ciliated microorganisms generate flows through flagellar or ciliary activity.
Even when the organism has little net translational motion, surface forcing can stir the surrounding fluid and alter scalar transport near the cell \cite{Goldstein2015GreenDynamics,Hartmann2007AnalysisAsymmetrica,Magar2003NutrientUptake,Liu2024FeedingEquivalent}.

Previous studies have shown that active suspensions can generate flows to enhance passive tracer dispersion \cite{Leptos2009DynamicsMicroorganisms, Thiffeault2010StirringBodies, Lin2011StirringSquirmers, Zaid2011LevyBacteria, Pushkin2013FluidMicroswimmers, Kanazawa2020LoopySuspensions, Jin2021CollectiveMicroswimmers, liu2026run}. 
For a single active cell, however, this enhancement is no longer homogeneous but spatially localized. Fluid velocities decay away from the organism \cite{Drescher2010DirectMicroorganisms, Drescher2011FluidScattering}, so tracer fluctuations are strongest near the surface and weaker farther away. The resulting transport cannot generally be represented by a single homogeneous effective diffusivity. Instead, active surface forcing creates a heterogeneous diffusion landscape surrounding the organism, analogous to space-dependent diffusivities derived for planar active carpets \cite{Mathijssen2018NutrientCarpets, Guzman-Lastra2021ActiveFluxes,  Aguayo2024FloatingEcosystems, Barros2025LayeredCarpets,  Barros2026ConfinedTransport}. However, these studies have not considered actively enhanced transport for finite-sized spherical geometries, such as the phycosphere.

In this paper, we investigate the biophysical connection between these ecological and transport processes by determining how activity-enhanced diffusion modifies solute exchange in the phycosphere [Fig.~\ref{fig:1}]. 
We first compute the space-dependent diffusivity generated by fluctuating flows and validate our predictions using stochastic simulations. 
We then solve the transport problems of nutrient uptake and metabolite release around an active cell. 
Finally, we couple the inward and outward carbon fluxes through metabolism, representing a simple photosynthetic turnover model in which enhanced carbon uptake supports enhanced release of oxygen and organic metabolites. 
This work shows that active mixing can both increase carbon acquisition and expand the radial region over which phytoplankton modify their chemical environment.

\section{Results}

Before computing the active fluxes generated by a planktonic microorganism, we first consider the passive flux set by molecular diffusion alone.
For simplicity, we consider an organism that is spherical with radius $a$, non-motile, and located in a quiescent fluid at the origin.
By solving Fick's laws in spherical coordinates, the carbon influx is given by
    \begin{equation}
    \label{eq:ThermalFlux}
        \Phi_\textmd{th} = 4\pi a C_\infty D_\textmd{th},
    \end{equation}
where $C_\infty$ is the background concentration of dissolved $\text{CO}_2$ in the ocean, 
$D_\textmd{th}$ is the thermal diffusivity, and we assume a perfectly absorbing boundary condition at the surface of the organism, $C(r=a)=0$, to obtain the largest possible flux.
This quantity scales linearly with the organism size $a$, whereas its metabolic demand typically scales with the volume, as $\Phi_\textmd{meta}=\beta a^3$.
Therefore, organisms with a radius larger than $a \sim \sqrt{4\pi C_\infty D_\textmd{th}/\beta}$ cannot rely on thermal diffusion alone.

\subsection{Hydrodynamic origin of enhanced diffusive transport}

To increase its carbon influx, the organism can actively stir its surrounding fluid by producing flows $\vec{u}(\vec{r})$ at low Reynolds number \cite{Lauga2009TheMicroorganisms, Elgeti2015PhysicsReview}, for example with cilia or flagella \cite{Bruot2016RealizingColloids, Gilpin2020TheFlagella, Wadhwa2022BacterialMechanisms}.
These viscosity-dominated flows can be decomposed using a multipole expansion of the Stokeslet, the fundamental solution of the Stokes equations \cite[see e.g.][]{Lauga2020TheMotility}.
Because the organism cannot exert a net force or torque on the fluid, this multipole expansion cannot contain monopoles (stokeslets and rotlets).
Therefore, to leading order, we model the organism flow as a Stokes dipole (stresslet), given by
    \begin{equation}
    \label{eq:dipole_flow_field}
        \vec{u}(\vec{r},t)=-\frac{3\mathcal P}{8\pi\mu r^2}\left(\cos^2\theta(t)-\frac 1 3\right) \vec{\hat{r}},
    \end{equation}
where $\mathcal P$ is the dipole strength, $\mu$ is the fluid viscosity, and $\theta(t)$ is the angle between the dipole axis and the position $\vec{r}$ where the flow is measured \cite{Lauga2020TheMotility}.
This angle fluctuates continuously, as the dipole orientation $\vec{\hat e}$ performs rotational diffusion on the unit sphere according to 
    \begin{equation}
    \frac{d\vec{\hat{e}}}{dt}=\vec\eta\times\vec{\hat{e}},\quad\langle \eta_i(t)\eta_j(t')\rangle=2D_r\delta_{ij}\delta(t-t'),
    \end{equation}
where $D_r$ is the rotational diffusivity, $\delta_{ij}$ is the Kronecker delta, and $\delta(t)$ is the Dirac delta function.

\begin{figure*}[t]
    \centering
    \includegraphics[width=1\linewidth]{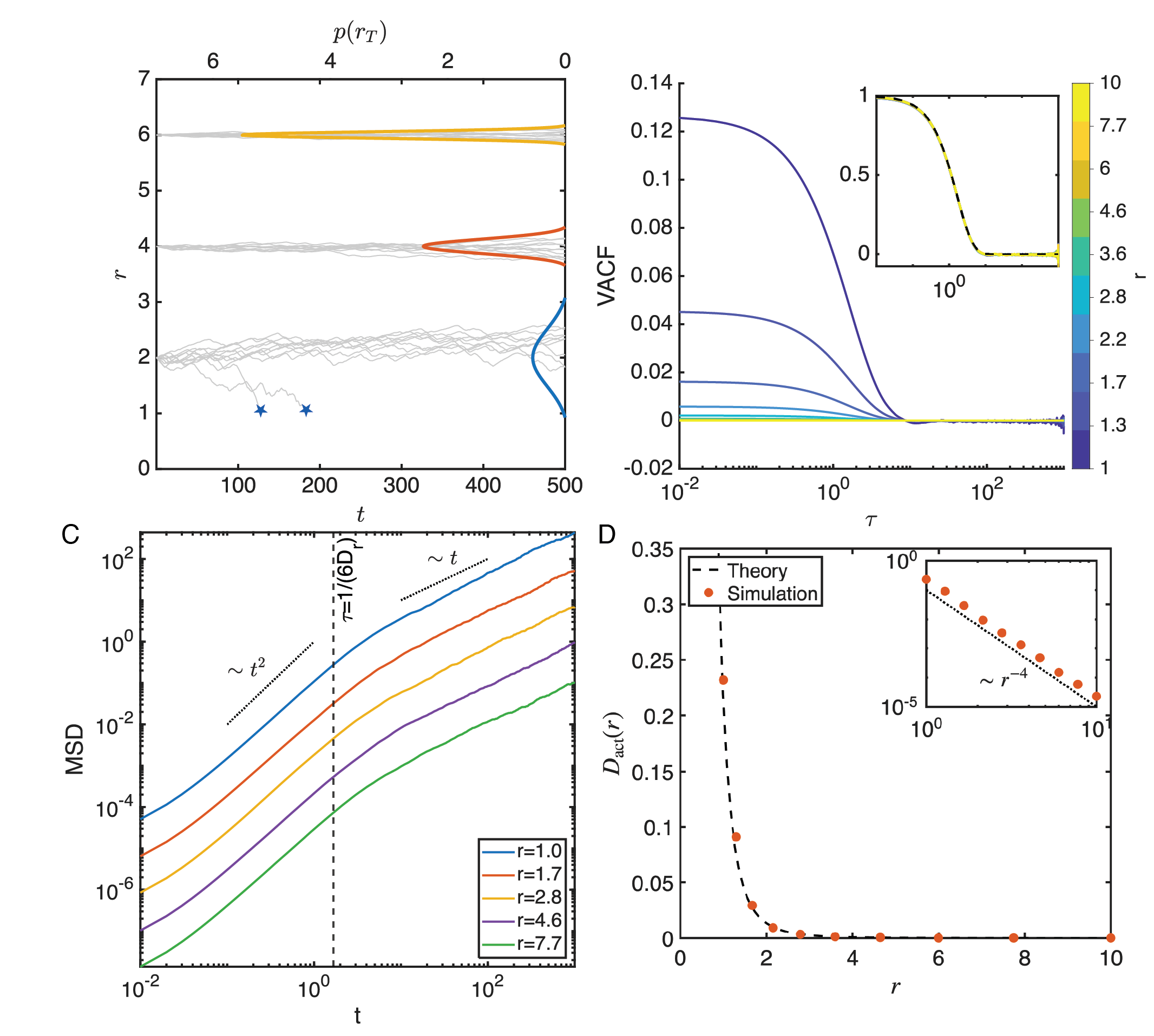}
    \caption{
        \label{fig:2}
        \textbf{Activity-induced diffusion landscape.}
    (A) Representative radial trajectories $r(t)$ for tracers released at different initial distances from the cell. A tracer is captured when it reaches the cell surface at $r=1$; star markers indicate absorption events. The top axis shows the corresponding probability distributions $p(r_T)$ at a fixed final time, determined by the local position-dependent diffusivity.   
    (B) Local tracer velocity fluctuations measured at fixed radius, with color indicating the tracer position as shown in the colorbar. The correlation decays on the reorientation timescale set by rotational diffusion, and its amplitude decreases strongly with distance from the cell. \textit{Inset:} collapsed VACF curves normalized by the theoretical amplitude, $(4/45)\left(3\mathcal P/(8\pi\mu r^2)\right)^2$, confirming the theoretical prediction [Eq.~\ref{eq:CorrelationFunction}].
    (C) Mean squared displacement (MSD) of tracer particles for $D_r = 0.1, \mathcal P = 10$, and $\mu = 0.1$, shown for different radial distances from the cell. The MSD exhibits two regimes separated by the characteristic time $\tau = (6D_r)^{-1}$: a ballistic regime at short times and a diffusive regime at long times.
    (D) Effective diffusion coefficient $D_{\mathrm{act}}(r)$ extracted from the long-time slope of the MSD for Eulerian tracers. The dashed line shows the theoretical prediction. \textit{Inset} log–log plot demonstrating the asymptotic decay $D_{\mathrm{act}} \sim r^{-4}$ [Eq.~\ref{eq:ActiveDiffusivity}].
    }
\end{figure*}

Next, we consider how these actively generated flows can mix the fluid by calculating the diffusivity of tracer particles as a function of distance from the organism.
Naturally, tracers close to the organism experience stronger flow fluctuations, while tracers far away experience weaker fluctuation. 
This is depicted in Fig.~\ref{fig:2}A, which shows how tracer trajectories evolve over time. 
Farther away, the tracer distribution spreads less [yellow], and nearer they spread more [blue]. 
This is the central idea: Actively generated flows create a spatially varying transport landscape.

To quantify these fluctuations, we first compute the orientational decorrelation of the stresslet. The probability distribution $P(\hat{\vec e},t)$ on the unit sphere satisfies the rotational diffusion equation
\begin{equation}
    \frac{\partial P}{\partial t}=D_r\nabla^2 P,
\end{equation}
where $\nabla^2$ is the angular Laplacian in spherical coordinates. Since the angular dependence of the stresslet is proportional to the $\ell=2$ spherical harmonic \cite{Berne1976DynamicPhysics}, we can derive that the autocorrelation function of the flow velocity is 
\begin{align}
    \langle u_r(t) u_r(0)\rangle
    \label{eq:CorrelationFunction}
    &=\left(\frac{3\mathcal P}{8\pi\mu r^2}\right)^2\frac{4}{45}\cdot e^{-6D_rt},
\end{align}
where the decorrelation timescale $\tau = (6D_r)^{-1}$.
This prediction is confirmed with numerical simulations in Fig.~\ref{fig:2}B. The correlation function indeed decays with timescale $(6D_r)^{-1}$, and its amplitude decreases as $1/r^4$ with distance from the cell.

The same timescale controls the tracer mean-squared displacement (MSD). At short times, $t\ll \tau$, the dipole orientation has not changed appreciably, so the tracer is advected by the local flow velocity and the motion is ballistic, according to
\begin{equation}
    \mathrm{MSD}(t)\approx \langle u_r^2\rangle t^2.
\end{equation}
At long times, $t\gg (6D_r)^{-1}$, the motion crosses over to an effective diffusive regime,
\begin{equation}
\mathrm{MSD}(t)\approx 2D_{\mathrm{act}}(r)t,
\end{equation}
as shown in Fig.~\ref{fig:2}C.

Using the Green-Kubo relation, and assuming that tracer displacements over one correlation time remain small compared with the scale on which the flow varies, the activity-induced diffusivity is
\begin{align}
\label{eq:dif_coef}
    D_{\mathrm{act}}(r)
    &=
    \int_0^\infty \langle u_r(t)u_r(0)\rangle\,dt \\
    \label{eq:ActiveDiffusivity}
    &=
    \left(\frac{3\mathcal P}{8\pi\mu r^2}\right)^2
    \frac{4}{45}\frac{1}{6D_r}
    = \frac{\mathcal P^2}{480\pi^2\mu^2D_r\,r^4}.
\end{align}
Thus, the induced diffusivity also decays as $r^{-4}$. This scaling is physically intuitive, as the stresslet velocity decay as $r^{-2}$, so the diffusivity scales as and $D\sim u^2\tau$. In the scalar transport problem below, we use this result as the radial component of an effective, spherically averaged diffusivity. Figure~\ref{fig:2}D confirms this prediction and shows excellent agreement between theory and simulation.

Although we focus on the stresslet case, the same logic applies more generally to fluctuating multipolar flows with algebraic spatial decay. 
To keep the analysis general, we therefore write the total diffusivity as
\begin{equation}
D(r)=D_\textmd{th}+\Gamma r^{-\alpha},
\end{equation}
where $D_\textmd{th} $ is the background thermal diffusivity and $\Gamma r^{-\alpha}$ is the activity-induced contribution. For the stresslet considered here, Eq.~\ref{eq:ActiveDiffusivity} gives $\alpha=4$ and
\begin{equation}
\Gamma=\frac{\mathcal P^2}{480\pi^2\mu^2D_r}.
\end{equation}
For higher-order multipoles, the exponent will be larger: $\alpha=6$ for quadrupoles that decay as $1/r^3$, and so on. 

We now turn to the consequences of this space-dependent diffusivity for nutrient uptake and scalar transport around the cell.

\subsection{Transport around an active cell}

We model the $\text{CO}_2$ (or nutrient) concentration field $c(r)$ outside the microorganism by the radial diffusion equation, which can be written as
\begin{equation}
    \nabla\cdot(D(r)\nabla c)=0,\quad D(r)=D_\textmd{th}+\frac\Gamma{r^\alpha},
\end{equation} 
where $D_\textmd{th}$ is the baseline diffusivity due to thermal fluctuations and $\Gamma r^{-\alpha}$ is the activity-induced contribution, which we derived above.
We first consider the problem of nutrient uptake by an absorbing cell. The concentration is fixed to zero at the cell surface, while the far-field is maintained at a constant value, corresponding to the boundary conditions $c(r=a)=0$ and $c(r=\infty)=C_\infty$. 
Solving the radial diffusion equation gives the steady-state concentration profile
\begin{equation}
    c(r)=C_\infty\left(1-\frac{\int_r^\infty\frac{ds}{D_\textmd{th} s^2 + \Gamma s^{2-\alpha}}}{\int_a^\infty\frac{ds}{D_\textmd{th} s^2 + \Gamma s^{2-\alpha}}}\right),
\end{equation}
and the total $\text{CO}_2$ influx is
\begin{equation}
    \label{eq:FluxLaw}
    \Phi=\frac{4\pi C_\infty}{{\int_a^\infty\frac{ds}{D_\textmd{th} s^2 + \Gamma s^{2-\alpha}}}}.
\end{equation}
Thus, the heterogeneous diffusivity modifies both the concentration profile and the uptake rate relative to purely passive diffusion. 

\begin{figure*}[t]
    \centering
    \includegraphics[width=0.7\linewidth]{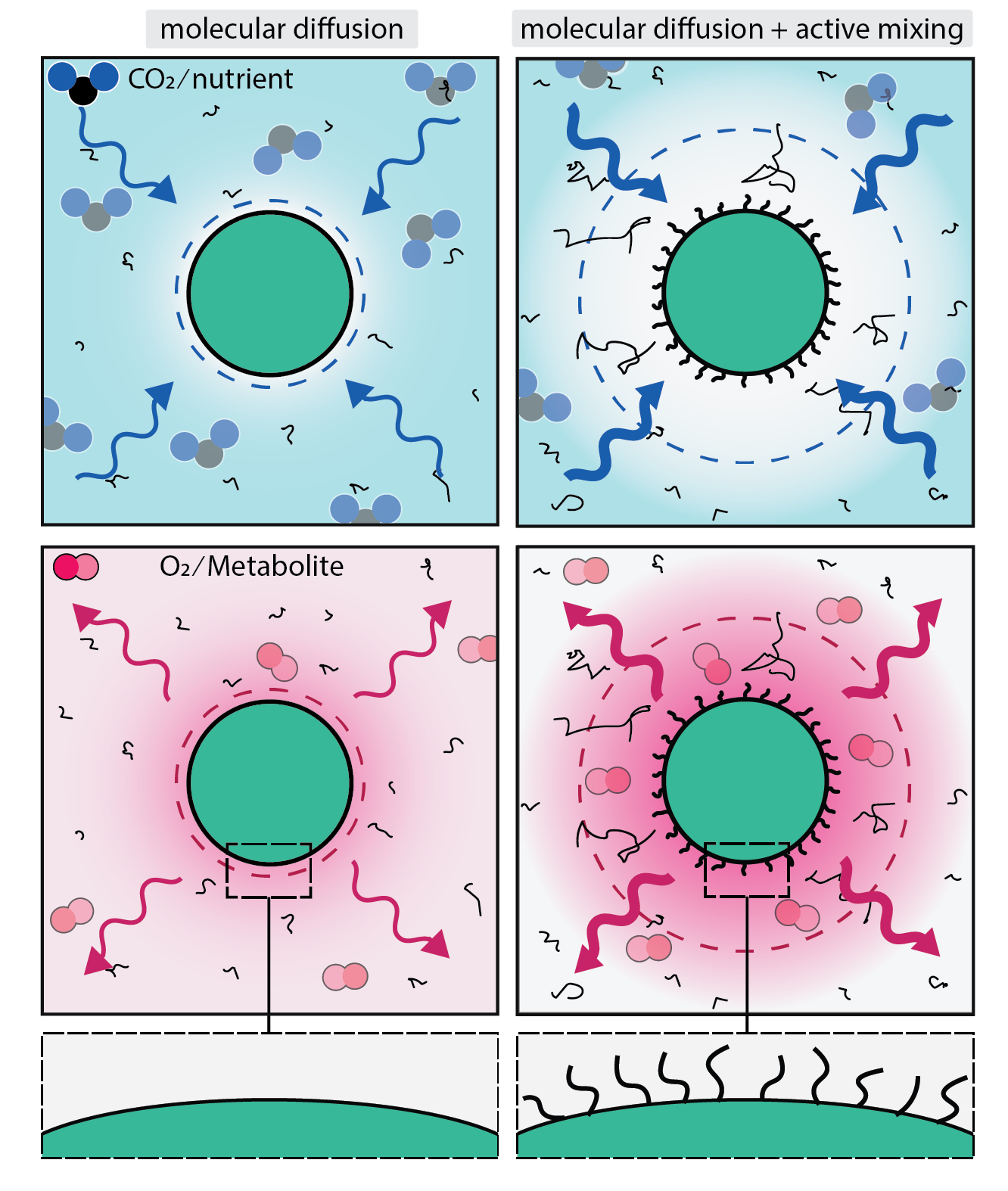}
    \caption{
        \label{fig:1}
        \textbf{Active surface forcing creates a heterogeneous transport region.}
    Schematic comparing purely thermal diffusion (left column) with active mixing plus thermal diffusion (right column). Top row: $\mathrm{CO}_2$ or nutrients are transported toward an absorbing cell. Bottom row: $\mathrm{O}_2$ or metabolites are released from the cell. In the passive case, transport is set by molecular diffusion around the physical cell radius. In the active case, surface forcing stirs the surrounding fluid, increasing the local effective diffusivity near the cell and broadening the region over which solutes are redistributed (dashed circle). Blue arrows indicate inward nutrient transport, pink arrows indicate outward metabolite transport, and background shading denotes the corresponding concentration fields. Insets in the bottom row illustrate the difference between a passive surface and an actively forced surface. 
    }
\end{figure*}

\begin{figure*}[t]
    \centering
    \includegraphics[width=1\linewidth]{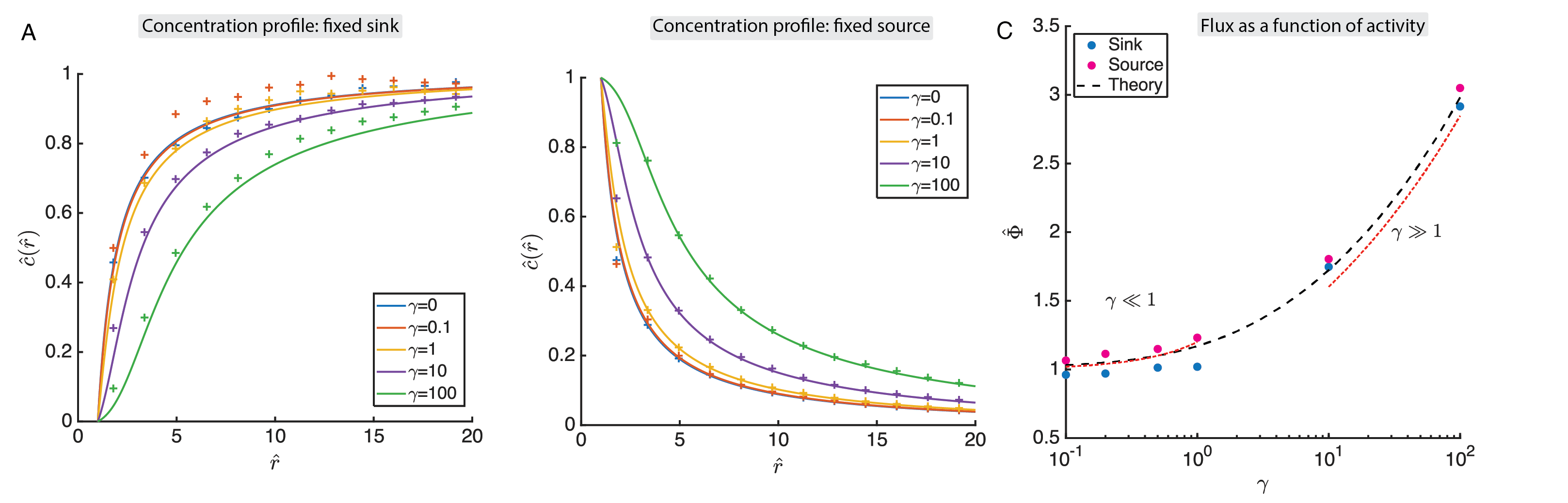}
    \caption{
        \label{fig:3}
        \textbf{Steady state concentration profiles and flux.}
    (A) Steady-state concentration profile $\hat{c}(\hat{r})$ for different activity strength $\gamma$ for cell as a sink. Nutrient concentration is fixed at $\hat r=\infty$. Solid lines denote theoretical predictions, while markers denote simulations. As $\gamma$ increases, the concentration profile broadens and is displaced farther from the cell, indicating that active mixing enhances effective transport and enlarges the region over which the cell perturbs its chemical environment.
    (B)  Steady-state concentration profile $\hat{c}(\hat{r})$ for different activity strength $\gamma$ for cell as a source. Metabolite concentration is fixed at the cell surface $\hat r=1$.
    (C) Flux enhancement as a function of activity strength $\gamma$, showing the universal crossover from weakly perturbed diffusion to activity-dominated transport. Colored symbols denote numerical results for sink and source boundary conditions, while the black dashed curve shows the full analytical prediction. Red dashed curves indicate the asymptotic limits for weak activity ($\gamma \ll 1$) and strong activity ($\gamma \gg 1$). In the low-activity regime, the flux remains close to its passive value, with only a small correction due to actively induced mixing. As $\gamma$ increases, the flux departs from the passive baseline and enters a strongly enhanced regime in which activity substantially enlarges the effective transport radius of the cell.     
    }
\end{figure*}

We verify this solution by simulating tracer particles subjected to the same space-dependent diffusivity [see \S\ref{subsec:sim_detail} and \S\ref{subsec:SDE} for details]. As shown in Fig.~\ref{fig:3}A, the simulated concentration profiles agree well with the analytical prediction. Increasing activity shifts the concentration profile outward [also see Fig.~\ref{fig:1}, top panels], reflecting enhanced transport near the cell and a corresponding increase in uptake.

The same framework also applies to a complementary ``source'' problem, where the organism excretes $\text{O}_2$ or metabolic products.
The solution follows directly from reversing the boundary values: The concentration of metabolites is set to a non-zero constant value at the cell surface and set to zero at infinity.
The derivation is discussed in \S\ref{subsubsec:fixed_surf_con}. 
The resulting concentration profile is depicted in Fig.~\ref{fig:3}B. Similar to the sink problem, an increase in activity broadens the concentration profile [also see Fig.~\ref{fig:1}, bottom panels].

\subsection{Asymptotic flux laws}
\label{subsec:asymp_flux}

Before we compute the fluxes [Eq.~\ref{eq:FluxLaw}], we first  identify the relevant control parameters by introducing the following dimensionless variables:
\begin{align}
    \hat r &= \frac{r}{a}, \qquad
    \hat t = \frac{D_\textmd{th}}{a^2}t, \\
    \hat c(\hat r) &= \frac{c(r)}{C_\infty}, \qquad 
    \hat\Phi = \frac{\Phi}{4\pi aD_\textmd{th}C_\infty}.
\end{align}
The dimensionless diffusivity is then
\begin{equation}
    \hat D(\hat r)=\frac{D(r)}{D_\textmd{th}}
    =1+\gamma \hat r^{-\alpha}, 
    \qquad
    \gamma\equiv \frac{\Gamma}{D_\textmd{th} a^\alpha},
\end{equation}
where $\gamma$ is the dimensionless activity parameter.
Physically, this quantity measures the strength of active diffusion relative to thermal diffusion at the cell scale. 
Finally, by defining the integral function
\begin{equation}
    \label{eq:J_int}
    J(\hat r;\gamma)=\int_{\hat r}^\infty \frac{d\xi}{\xi^2(1+\gamma\xi^{-\alpha})},
\end{equation}
we find the dimensionless flux
\begin{equation}
    \label{eq:DimensionlessFlux}
    \hat\Phi(\gamma)=\frac{1}{J(1,\gamma)}.
\end{equation}

This solution is shown in Fig.~\ref{fig:3}C for different degrees of activity.
This analytical prediction agrees well with our numerical results for all $\gamma$ values. 
By symmetry, the source and sink problems have the same flux magnitude and differ only in the direction of transport.

In the weak activity limit, $\gamma\ll 1$, expanding the integral function [Eq.~\ref{eq:J_int}] yields the flux
\begin{equation}
    \hat\Phi(\gamma)\approx 1+\frac{\gamma}{1+\alpha}.
\end{equation}
In this case, the uptake enhancement grows linearly with $\gamma$. 
The prefactor decreases with increasing $\alpha$, so higher multipoles contribute less.
This is consistent with our earlier analysis that rapidly decaying flow fields perturb only a narrow region near the cell.

In the strong-activity limit, $\gamma\gg 1$, expanding the integral gives
\begin{equation}
\hat\Phi(\gamma)\approx \frac{\alpha\sin(\pi/\alpha)}{\pi}\gamma^{1/\alpha}.
\end{equation}
The uptake flux then grows sub-linearly with activity. Even a very strong stresslet cannot enhance uptake unboundedly, because transport remains limited by the outer diffusive region. In this case, additional efforts only provide diminishing returns.

Interestingly, the enhanced fluxes can be interpreted naturally using an effective transport radius, $a_\mathrm{eff}$.
For a passive cell, this radius is simply set by the physical radius $a$. For an active cell, however, activity enlarges the region over which solute is efficiently redistributed. 
We therefore define 
\begin{equation}
    a_\mathrm{eff}= \hat\Phi(\gamma) a,
\end{equation}
which is the radius a passive cell would need to achieve the same uptake flux.
In the weak-activity regime,
\begin{equation}
    a_\mathrm{eff}\approx a+\frac{\Gamma}{(1+\alpha)D_\textmd{th} a^{\alpha-1}},\quad\gamma\ll1,
\end{equation}
so activity acts as an additive correction to the physical cell size.
In the strong-activity regime,
\begin{equation}
    a_\mathrm{eff}\approx\alpha\frac{\sin(\pi/\alpha)}{\pi}\left(\frac{\Gamma}{D_\textmd{th}}\right)^{1/\alpha},\quad \gamma\gg1,
\end{equation}
the effective transport radius becomes independent of the physical radius $a$ and is instead controlled by the active mixing scale.



\subsection{Coupled uptake and release expands the phycosphere}

\begin{figure*}[t]
    \centering
    \includegraphics[width=1\linewidth]{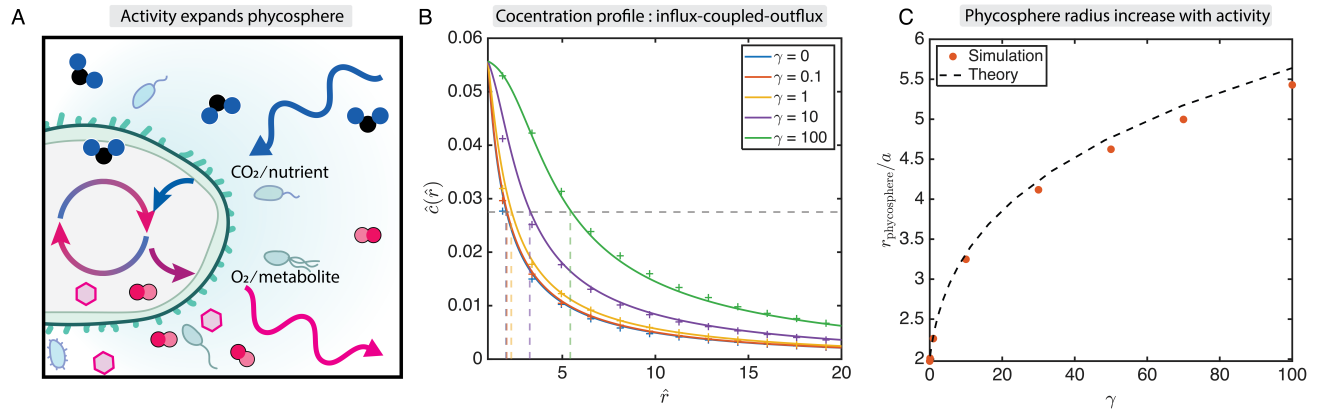}
    \caption{
        \label{fig:4}
        \textbf{Coupled uptake and release expands the phycosphere.}
   (A) Schematic of coupled nutrient uptake and metabolite release. Uptake of $\mathrm{CO}_2$ or nutrients supports metabolism, which releases $\mathrm{O}_2$ and metabolites into the surrounding fluid. If the outward flux is proportional to the nutrient influx, active mixing increases the total released flux and redistributes metabolites over a larger region.    
    (B) Concentration profiles of the tracers when $\Phi_\mathrm{out}=\kappa\Phi_\mathrm{in}$, 
    with $\kappa=0.7$. Solid lines show theory and markers show simulations for different activity strengths $\gamma$. The surface concentration remains fixed at $\kappa C_\infty$, while increasing activity broadens the profile. The horizontal dashed line marks half the surface concentration, and the vertical dashed lines mark the corresponding phycosphere radius, $\hat r_\mathrm{phycosphere}$. 
    (C) Phycosphere radius as a function of activity strength $\gamma$, defined by the radial position where the released concentration falls to one half of its surface value. The radius approaches the passive value at weak activity and grows as $\sim\gamma^{1/\alpha}$ in the strong-activity regime, matching the scaling of the activity-enhanced transport radius. 
}
\end{figure*}

Until now, we considered the influx and outflux as independent transport problems. 
However, uptake of $\mathrm{CO}_2$ and nutrients supports cellular metabolism, which in turn drives the release of $\mathrm{O}_2$, dissolved organic carbon, and other metabolites. 
We model this coupling by prescribing the total outflux as a fraction of the influx,
\begin{equation}
    \Phi_\mathrm{out}=\kappa\Phi_\mathrm{in},
\end{equation}
where $\kappa$ is the constant of proportionality.
This simple model captures the idea that enhanced carbon delivery can further support the production and release metabolites, as sketched in Fig.~\ref{fig:4}A.

Using the fixed-flux source solution from \S\ref{subsubsec:const_surf_flux} together with the uptake flux computed above [Eq.~\ref{eq:DimensionlessFlux}] gives
\begin{equation}
c_\mathrm{out}(\hat r)=\kappa C_\infty\frac{J(\hat r;\gamma)}{J(1;\gamma)}.
\end{equation}
At the surface, $\hat r=1$, this gives $c_\mathrm{out}(1)=\kappa C_\infty$. Thus, when metabolite production is proportional to nutrient uptake, activity increases the total outward flux without changing the surface concentration. Its main effect is to broaden the released chemical field.

We quantify this broadening by defining a phycosphere radius as the distance at which the released concentration has fallen to half its surface value:
\begin{equation}
c_\mathrm{out}(\hat r_\textmd{phycosphere})=\frac{1}{2}c_\mathrm{out}(1).
\end{equation}
In the weak-activity limit, the passive profile is recovered and $\hat r_\textmd{phycosphere}=2+O(\gamma)$. In the strong-activity limit, setting $\hat r=\gamma^{1/\alpha}y$ gives
\begin{equation}
    \hat r_\textmd{phycosphere}\sim y_{1/2}\gamma^{1/\alpha},
\end{equation}
where $y_{1/2}$ is determined by
\begin{equation}
    \int_{y_{1/2}}^\infty \frac{dy}{y^2+y^{2-\alpha}}
    =\frac{1}{2}\int_0^\infty \frac{dy}{y^2+y^{2-\alpha}}.
\end{equation}
Thus, in the coupled uptake-release model, activity expands the phycosphere through two linked effects. First, it increases the nutrient influx that sets the total metabolite release rate. Secondly, it redistributes the released metabolites over a larger radial region. The half-concentration radius therefore grows from the passive value, $\hat r_{\mathrm{phycosphere}}=2+O(\gamma)$, to an activity-dominated length scale proportional to $\gamma^{1/\alpha}$. This scaling is the same as the active transport radius obtained from the flux enhancement, showing that increased uptake and expanded chemical signaling are controlled by a common hydrodynamic mixing scale. Figure \ref{fig:4}B shows the resulting concentration profiles at the various activity levels, and figure \ref{fig:4}C summarizes this increase of $\hat r_{\mathrm{phycosphere}}$ with activity.


\section{Discussion}

Our results show that active surface forcing can enlarge the phycosphere by reshaping solute transport around phytoplankton. Rather than acting only as passive absorbers or secretors, hydrodynamically active cells can modify the effective radius over which they acquire carbon and nutrients and release oxygen or metabolites. This provides a physical mechanism by which ciliary or flagellar activity may influence both phytoplankton physiology and the chemical landscape experienced by nearby bacteria.

This mechanism is directly relevant to bacteria-phytoplankton interactions in the phycosphere. Marine bacteria often respond chemotactically to phytoplankton-derived dissolved organic matter, and the spatial distribution of these exudates controls encounter rates, colonization, and remineralization. \cite{Smriga2016ChemotaxisBacteria, Stocker2008RapidPatches} In our model, active mixing broadens the metabolite field without necessarily changing the surface concentration when release is coupled to uptake. Thus, activity can increase the volume over which bacteria experience phytoplankton-derived chemical cues. At the same time, enhanced inward transport can increase the supply of $\mathrm{CO}_2$ or nutrients to the phytoplankton surface, linking microscale hydrodynamics to cellular carbon fixation. This broadened chemical field may also connect phycosphere-scale transport to marine snow formation, since phytoplankton-derived organic matter and bacterial colonization are key steps in the development of particle-associated microbial communities \cite{Buchan2014MasterRecyclers,Passow2002TransparentEnvironments,Azam2001SeaMicrocosms,Simon2002MicrobialEcosystems}.

The activity considered here need not rely solely on the phytoplankton. Bacteria recruited to the phycosphere generate their own flagellar flows, and dense bacterial suspensions can produce collective fluid motion \cite{Drescher2011FluidScattering,lushi2014}. These bacterial flows could add to the local mixing \cite{Mathijssen2018NutrientCarpets}, increasing $\mathrm{CO_2}$ or nutrient influx to the phytoplankton. If the resulting increase in uptake drives greater release of $\mathrm{O_2}$ or other metabolites as postulated here, the stronger chemical field could in turn recruit more chemotactic bacteria \cite{Stocker2012MarineGradients, Stocker2012}. This suggests a possible positive feedback loop in which bacterial accumulation, hydrodynamic mixing, phytoplankton uptake, and metabolite release reinforce one another. Future models could capture this effect by allowing the activity parameter to depend on bacterial abundance and chemotactic recruitment. 

The model also has clear limits. We used a single fluctuating stresslet, assumed radial symmetry, and neglected mean advective transport, biochemical reactions, nutrient saturation, and feedback between chemical fields and microbial motion. These approximations isolate the role of heterogeneous diffusion, but future work could relax them by considering other hydrodynamic singularities, finite correlation times beyond rotational diffusion, reactive uptake, or collections of nearby active cells with overlapping transport landscapes.

Finally, estimating $\gamma$ from experimentally measured dipole strengths, rotational diffusivities, and molecular diffusivities will clarify where real microorganisms lie within the theoretical phase space. Preliminary estimates suggest that biologically relevant systems may lie near the crossover between weak and strong activity, making the nonlinear transport enhancement predicted here potentially observable.

\section*{Acknowledgments}

We thank all members of the Mathijssen lab for their support and insightful discussions. A.J.T.M.M. acknowledges funding from the Charles E. Kaufman Foundation (Early Investigator Research Award KA2022-129523; and New Initiative Research Award KA2024-144001), the National Science Foundation (Career Award CBET-2542731; and UPenn MRSEC DMR-2309043), the University of Pennsylvania (URF, CURF, VIPER, Vagelos MLS, and FERBS programs), and the Research Corporation for Science Advancement (Cottrell Scholar Award CS-CSA-2026-125).

\section*{Author contributions}
A.M. designed the research and provided funding. M.L. developed the analytical theory, performed the simulations, and made the figures. Both authors analyzed the data and wrote the manuscript.

\section*{Competing interests}
The authors declare no competing interests.

\section*{Data availability}
The data supporting the findings of this article are openly available at \cite{LiuData2026}.

\section*{Code availability}
The simulation code used in this paper is available from \url{https://github.com/yanagi814/heterogenous_diffusion}

\section{Supplementary material}
\subsection{Cell as a source}
\subsubsection{Fixed surface concentration}
\label{subsubsec:fixed_surf_con}
The same transport framework can also be used to describe the dispersion of a scalar released from the cell surface. As a complementary problem, we consider a fixed-concentration source, with boundary conditions $c(a)=C_a,\,c(\infty)=0,$ corresponding to a cell surface maintained at concentration $C_\infty$ and a vanishing far-field concentration. 
A schematic comparison between the active vs passive settings is shown in Fig.~\ref{fig:3}A.
Solving the radial diffusion equation yields
\begin{equation}
    c_{\mathrm{source}}(r)=C_\infty\left(1-
    \frac{\int_a^r \frac{ds}{D_\textmd{th}s^2+\Gamma s^{2-\alpha}}}
         {\int_a^\infty \frac{ds}{D_\textmd{th}s^2+\Gamma s^{2-\alpha}}}
    \right).
\end{equation}
The corresponding outward flux is
\begin{equation}
    \Phi_{\mathrm{source}}=-\frac{4\pi C_\infty}{\int_a^\infty \frac{ds}{D_\textmd{th}s^2+\Gamma s^{2-\alpha}}},
\end{equation}
which differs from the uptake case only by a sign, reflecting the reversal of transport direction.


Thus, the same heterogeneous diffusivity that enhances nutrient uptake also enhances outward transport away from the cell. 
\subsubsection{Constant surface flux}
\label{subsubsec:const_surf_flux}
Instead of prescribing the concentration at the cell surface, another model for metabolic waste release is to prescribe the secretion flux. Let the outward flux density at the cell be $w$. The total secretion rate is $\Phi_\mathrm{out}=4\pi a^2 w$. 

Solving the same radial diffusion equation with boundary conditions $-D(a)\frac{dc}{dr}\Big|_{r=a}=w,\, c(\infty)=0.$ gives the concentration profile 
\begin{equation}
    c(r)=\frac{W}{4\pi}\int_r^\infty \frac{ds}{D_\textmd{th}s^2+\Gamma s^{2-\alpha}}.
\end{equation}

Unlike the fixed-concentration source problem, in which the flux is determined by the transport field, the total outward flux is here prescribed by the secretion rate $W$. Activity therefore does not alter the total flux itself, but instead changes the concentration profile required to sustain that flux. Increasing activity reduces the concentration buildup near the cell, thereby enhancing waste .

\subsection{Simulation details}
\label{subsec:sim_detail}
The concentration profiles and fluxes in Fig.~\ref{fig:3} were computed using MATLAB. Unless otherwise noted, the dimensionless parameters were $a=1$, $R=80$, $D_\textmd{th}=1$, $\alpha=4$, and $\lambda=0$. Dirichlet boundary condition simulations used $N_p=5000$ particles with time step $\Delta t=\min(10^{-2},1/(100\gamma))$, burn-in time $T_{\mathrm{burn}}=1000$, and sampling time $T_{\mathrm{sample}}=1000$.

The computational domain is a spherical shell, $a\le r\le R$, with the inner and outer boundaries used as absorbing or reservoir boundaries depending on the boundary condition being simulated. Boundary-crossing events are counted to estimate the flux, and particles are reinjected near the appropriate reservoir boundary to maintain the imposed concentration. Reinjection is performed within a thin shell rather than at a single radius, which reduces artificial boundary recrossing.

After the burn-in period, particle positions are sampled over time and binned into radial shells. The concentration in shell $j$ is estimated from the time-averaged particle number divided by the shell volume,
\begin{equation}
    c_j\propto \frac{\langle N_j\rangle_t}{(4\pi/3)(r_{j+1}^3-r_j^3)}.
\end{equation}
The flux is estimated independently by counting absorption events during the sampling window and normalizing by the sampling time and the imposed reservoir concentration. The analytical curves plotted with the simulations are evaluated using the same finite outer radius $R$ as the simulations.

\subsection{SDE formulation}
\label{subsec:SDE}
The particle simulations sample the diffusion equation with space-dependent diffusivity,
\begin{equation}
    \frac{\partial c}{\partial t}=\nabla\cdot\left(D(\mathbf x)\nabla c\right),
    \qquad
    D(r)=D_\textmd{th}+\Gamma r^{-\alpha}.
\end{equation}
For multiplicative noise, the stochastic update depends on the convention used to evaluate the diffusivity during a time step. $\lambda=0$ corresponds to the It\^o convention, $\lambda=1/2$ to the Stratonovich convention, and $\lambda=1$ to the isothermal or H\"anggi--Klimontovich convention \cite{MeerschaertChapterEquations}.

We use the explicit It\^o convention, $\lambda=0$, because it can be integrated directly with Euler--Maruyama. To recover the desired Fickian diffusion equation, the It\^o update must include the noise-induced drift associated with the diffusivity gradient \cite{Vaccario2015First-PassageMedia}. The simulated SDE is
\begin{equation}
    d\mathbf X_t = \nabla D(\mathbf X_t)\,dt
    +\sqrt{2D(\mathbf X_t)}\,d\mathbf W_t ,
\end{equation}
which is advanced numerically as
\begin{align}
    \mathbf X_{n+1}
    &=\mathbf X_n+\nabla D(\mathbf X_n)\Delta t
    +\sqrt{2D(\mathbf X_n)\Delta t}\,\boldsymbol\xi_n,\\
    \boldsymbol\xi_n&\sim\mathcal N(\mathbf 0,\mathbf I).
\end{align}

Thus, for the It\^o implementation used here, the drift is $\nabla D$. For the radial diffusivity above,
\begin{equation}
    \nabla D(r)=\frac{dD}{dr}\hat{\mathbf r}
    =-\alpha\Gamma r^{-\alpha-1}\hat{\mathbf r}.
\end{equation}

This drift is not an additional physical force. It is the correction required so that the stochastic particle ensemble obeys the same diffusion equation solved analytically\cite{Lau2007State-dependentFormulation}. Indeed, the Fokker--Planck equation of the It\^o SDE is
\begin{align}
    \frac{\partial c}{\partial t}
    &=-\nabla\cdot\left(c\nabla D\right)+\nabla^2(Dc)\\
    &=\nabla\cdot\left(D\nabla c\right).
\end{align}

\bibliography{references-2}

\newpage

\end{document}